\documentclass[twocolumn, amssymb, nobibnotes, aps, amsmath, amssymb, superscriptaddress,floatfix]{revtex4-2}
\usepackage{graphicx}
\usepackage{dcolumn}
\usepackage{bm}
\usepackage[linktocpage,colorlinks=true,linkcolor=blue,citecolor=blue,breaklinks=true,urlcolor=blue]{hyperref}
\usepackage[utf8]{inputenc}
\usepackage[T1]{fontenc}
\usepackage{amsmath}
\DeclareGraphicsExtensions{.pdf,.jpeg,.jpg, .png, .eps, .tiff}
\usepackage{epstopdf}
\usepackage{lipsum}
\usepackage{natbib}
\usepackage[dvipsnames]{xcolor}
\usepackage[normalem]{ulem}
\usepackage{soul}
\usepackage{enumitem}
\usepackage{siunitx}
\usepackage{subfigure}
\usepackage{float}
\usepackage{gensymb}
\AppendGraphicsExtensions{.tga}
\DeclareMathOperator\erfi{erfi}
\renewcommand{\vec}[1]{\underline{#1}}

\newcommand{\vecP}[2]{{#1}_{#2}}
\newcommand{\tens}[1]{\underline{\underline{#1}}}
\newcommand{\kbt}{k_\text{B}T}
\newcommand{\VCO}{\Xi}
 
\newcommand{\abs}[1]{\left| #1 \right|}

\newcommand{\av}[1]{\left\langle #1 \right\rangle}
\newcommand{\Brac}[1]{\left( #1 \right)}

\newcommand{\tr}[1]{\mathrm{tr}\left[ #1 \right] }

\newcommand{\fig}[1]{Fig.~\ref{#1}}
\newcommand{\FIG}[1]{Figure~\ref{#1}}
\newcommand{\eq}[1]{Eq.~\ref{#1}}

\newcommand{\sctn}[1]{\S~\ref{#1}}

\begin{document}
\title{Dynamics of polymers in coarse-grained nematic solvents}

\newcommand{\ue}{School of Physics and Astronomy, The University of Edinburgh, Peter Guthrie Tait Road, Edinburgh, EH9 3FD, United Kingdom}
\newcommand{\csu}{Department of Physics and Astronomy, California State University, Long Beach, Long Beach, California 90840, USA}
\newcommand{\lu}{School of Mathematics, Loughborough University, Leicestershire, LE11 3TU, United Kingdom}

\author{Zahra Valei}
\affiliation{\ue}
\author{Karolina Wamsler}
\affiliation{\ue} 
\author{Alex J. Parker}
\affiliation{\lu}
\author{Therese A. Obara}
\affiliation{\csu}
\author{Alexander R. Klotz}
\affiliation{\csu}
\author{Tyler N. Shendruk}
\email{t.shendruk@ed.ac.uk}
\affiliation{\ue}

\begin{abstract}
\noindent
    Polymers are a primary building block in many biomaterials, often interacting with anisotropic backgrounds. 
    While previous studies have considered polymer dynamics within nematic solvents, rarely are the the effects of anisotropic viscosity and polymer elongation differentiated. 
    Here, we study polymers embedded in nematic liquid crystals with isotropic viscosity via numerical simulations, to explicitly investigate the effect of nematicity on macromolecular conformation and how conformation alone can produce anisotropic dynamics.
    We employ a hybrid technique that captures nematic orientation, thermal fluctuations and hydrodynamic interactions.
    The coupling of the polymer backbone to the nematic field elongates the polymer, producing anisotropic diffusion even in nematic solvents with isotropic viscosity.
    For intermediate coupling, the competition between background anisotropy and macrmolecular entropy leads to hairpins --- sudden kinks along the backbone of the polymer.
    Experiments of DNA embedded in a solution of rod-like fd viruses qualitatively support the role of hairpins in establishing characteristic conformational features that govern polymer dynamics.
    Hairpin diffusion along the backbone exponentially slows as coupling increases.
    Better understanding two-way coupling between polymers and their surroundings could allow the creation of more biomimetic composite materials.
\end{abstract}
\maketitle


\section{Introduction}
Composite materials are ubiquitous in biology, with their versatile and functional macroscopic properties arising from greater complexity compared to single constituent counterparts~\cite{eder2018}.
Biopolymer composites, such as microtubules in filamentous actin~\cite{Kikuchi2009}, filamentous bacteriophages in pathogenic biofilms~\cite{secor2015filamentous}, polysaccharides components in cell walls~\cite{bidhendi2020fluorescence}, chiral chitin~\cite{oh2016} and mucus~\cite{bansil2018biology,witten2018selective,werlang2019}, exemplify mesoscale constituents suspended within already complex soft material backgrounds.
Of these, polymers embedded in liquid crystalline solvents (so-called hypercomplex liquid crystals~\cite{dogic2014}) are particularly interesting because they idealize the competition between the broken symmetry of the surrounding medium and the tendency of the suspended phase to maximize entropy.

Macromolecules in good, isotropic solvents possess many internal degrees of freedom, such that entropy maximization encourages them to adopt random-coil configurations on scales greater than their Kuhn length.
On the other hand, polymers suspended in liquid crystalline solvents can be highly extended along the nematic director~\cite{becerra2024conformational}, with fluctuations away from perfect alignment quantified by the orientation distribution of the main polymer axis~\cite{Barbara2007} and by the Odijk deflection length~\cite{odijk1986}.
In fact, not only do they align with the director, but semiflexible polymers are observed to possess enhanced orientational order compared even to the background liquid crystal.
This surprising result has been experimentally demonstrated by direct single-molecule visualizations of semiflexible F-actin filaments, worm-like micelles and neurofilaments suspended in nematic phase solutions of rod-like virus particles~\cite{dogic2004elongation}, as well as conjugated polymers in 5CB~\cite{lammi2004}.
Increasing contour length further increases the measured polymer orientational order~\cite{dogic2004elongation}, though increasing segmentation within conjugated polymers decreases order~\cite{link2005}. 
Accompanying these changes in conformations are changes to the dynamics.
Polymers in nematic surroundings exhibit anisotropic diffusion ~\cite{turiv2013effect,link2006}, but it is not immediately clear if the anisotropic diffusion is due to the fluid's anisotropic viscosity or if it arises because of the elongation of the polymer.

While the interactions between component molecules lead to nematic alignment and complex dynamics, they are fundamentally difficult to simulate because of the division of time scales between the nematic surroundings and the polymeric solute.
Indeed for this reason, numerical studies of the dynamics of semiflexible polymers within an ensemble of many nematically ordered chains are more common than simulations of single macromolecules within mesophase nematics~\cite{lowen1999,egorov2016}.
The profound lack of numerical techniques for efficiently simulating macromolecules in liquid crystalline solvents demands novel simulation techniques be utilized if we wish to fundamentally understand the physical principles that lead to the unusual mechanical properties of binary mixtures of semiflexible polymers and low-molecular-weight nematogens.

Motivated by single-molecule microscopy images of DNA suspended in a solution of nematically ordered fd viruses, we introduce a hybrid mesoscopic simulation method based on molecular dynamics (MD) and nematic multi-particle collision dynamics (N-MPCD).
This hybrid approach employs standard MD for simulating semiflexible macromolecules and coarse-grained N-MPCD for modelling the nematic fluid, including diffusion, director fluctuations and hydrodynamic interactions. 
We investigate the configurational dynamics of a single chain by examining the relationship between hairpins along the backbone of polymer and polymer configuration, as well as the diffusion of hairpins. 
We observe that the hairpins themselves diffuse along backbone of the polymer, while the polymer exhibits anisotropic diffusion even though the fluid viscosity is isotropic.  
Our simulations provide evidence that the macromolecular conformation can be just as significant as the anisotropy of the viscosity in governing the diffusional anisotropy ratio. 

\section{Experiments}
Here, we qualitatively examine the effects of a nematic solvent on DNA molecules that are approximately 1000 persistence lengths long.
We use T4 DNA (169 kbp) stained with YOYO-1 fluorescent dye, leading to a contour length of approximately 60 $\mathrm{\mu m}$, compared to a persistence length of approximately 50 nm.
The DNA is embedded in a solution of rod-like fd viruses, which are 880 nm long and 6.6 nm in diameter\footnote[3]{Nematic solutions were obtained from Christopher Ramirez and Zvonimir Dogic at the University of California, Santa Barbara.}.
The virus solution concentration is 20 $\mu$g/mL in an aqueous solvent with an ionic strength of 20 mM, placing them in the weakly cholesteric nematic phase ~\cite{Dogic1997}. 
DNA is stained in the same ionic conditions and mixed with the virus solution at a 4:1 ratio, leaving the viruses in a nematic phase despite the lower concentration. 
Nematicity is verified by viewing the solution under cross-polarized microscopy; it is birefringent but not iridescent, consistent with previous characterization of the cholesteric phase ~\cite{Dogic1997}. 

Because nematic fd virus solutions are a poor solvent for DNA ~\cite{dogic2004elongation}, the equilibrium configuration of the DNA molecules is a diffraction limited globule. 
However when shear is applied, the molecules are transiently elongated within the nematic. 
Molecules are sheared by confining a solution of fd-DNA between unsealed glass coverslips separated by approximately 50 $\mathrm{\mu m}$, and sliding one cover slip with respect to the other.

\begin{figure}
    \centering
    \includegraphics[width=0.99\linewidth]{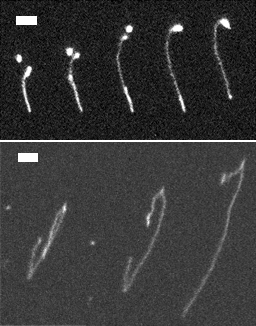}
    \caption{Hairpins along T4 DNA in nematic fd virus solution after shear. 
    (top) A DNA molecule with a $\approx180^\circ$ hairpin unfolding at the bottom. The bright object near the top is a second DNA molecule. 
    (bottom) A DNA molecule with two hairpins. The bottom hairpin unfolds. Scale bar in both images is 2 microns.}
    \label{fig:t4fd}
\end{figure}

Molecules are seen with sharp bends at acute angles (\fig{fig:t4fd}), qualitatively different from DNA sheared in isotropic fluids~\cite{tu2020direct}.
\FIG{fig:t4fd} shows two such examples, one molecule possessing an acute backfold (\fig{fig:t4fd}; top) and the other having two acute hairpins (\fig{fig:t4fd}; bottom).
Unfolding requires one arm of the hairpin to slide while maintaining the angle with the other.
This indicates that the DNA hairpins are topologically protected and can only be removed when the chain end reaches the joint (or two hairpins annihilate).
A complete experimental characterization of this system is a subject for future work, but these qualitative observations suggest that hairpins are characteristic features of polymers embedded in a nematically ordered background.
To investigate this assertion more quantitatively, we perform simulations of the suspension of flexible polymers in nematic liquid crystalline media. 

%

\section{Numerical model and Method}

To quantitatively study the conformations and dynamics of polymers embedded in nematic background, we employ a hybrid approach of N-MPCD to simulate the nematic background and MD to simulate the polymeric inclusion.
N-MPCD fully incorporates thermal fluctuations, hydrodynamic interactions and nematic orientational order of the liquid crystal (\sctn{sec:N-MPCD}).
MD discretizes the polymer into a linear sequence of bound beads that exchange momentum with fluid particles (\sctn{sec:MD}).
Additionally, polymer segments are coupled to the nematic orientation via a two-way coupling mechanism.

\subsection{Nematohydrodynamic Model}
\label{sec:N-MPCD}
While many microscopic particle-based methods explicitly calculate effective pair potentials between mesogen molecules~\cite{care2005,brini2013,zannoni2018,allen2019}, mesoscopic models further abstract the interactions between mesogens.
An early example is the Lebwohl-Lasher model, which models the nematic phase using nematogens fixed on a cubic lattice, with interactions between particles governed by pair potentials~\cite{lebwohlLasher1972,das2017}.
Since then other mesoscale algorithms have been conceived to simulate liquid crystal hydrodynamics, including multi-particle collision dynamics (N-MPCD) schemes~\cite{shendruk2015}.
N-MPCD employs coarse-grained collision operators to evolve the density, velocity, and orientation fields. 
Since the N-MPCD algorithm discretizes the nematic fluid into point-particles that interact through a stochastic, many-particle collision operator, the simulation time is reduced compared to methods that calculate the pair interactions.
We build on N-MPCD's success in simulating electroconvection~\cite{lee2017} and colloidal liquid crystals~\cite{reyes2020defects,wamsler2024lock,hijar2020dynamics}to consider polymer dynamics in liquid crystalline solvents.

The nematic fluid is discretized into point particles labeled $i$.
Each particle possesses a mass $m_i$, position $\vecP{\vec r}{i}\Brac{t}$, velocity $\vecP{\vec v}{i}\Brac{t}$ and nematic orientation $\vecP{\vec u}{i}\Brac{t}$~\cite{shendruk2015}.
The number of lines under the notation indicates the tensor rank, scalars are rank-0 tensors, vectors are rank-1 tensors, and so forth.
While time is discretized into discrete time steps $\delta t$, the other quantities evolve continuously.
MPCD is composed of two steps: $\Brac{i}$ a streaming step and $\Brac{ii}$ a collision step. 
In the streaming step, particle positions translate ballistically according to $\vecP{\vec r}{i}\Brac{t+\delta t}=\vecP{\vec r}{i}\Brac{t}+\vecP{\vec v}{i}\Brac{t}\delta t$. 
In the collision step, particle velocities update by binning particles into cells (labeled $c$) of size $a$.
Within each cell momentum is conserved but stochastically exchanged between particles through a collision operator characterized by the thermal energy $\kbt$~\cite{Noguchi2007}.
In this study, we use the Andersen-thermostatted version of the MPCD algorithm.
Orientations are modified by a second stochastic multi-particle collision operator, which reproduces the canonical distribution of the Maier-Saupe mean-field approximation about the local director $\vecP{\vec n}{c}$ within each MPCD cell~\cite{shendruk2015}.
The orientation collision operator is characterized by a globally specified nematic interaction constant $U_\text{NI}$. 
Velocity-orientation coupling is achieved by application of Jeffery theory for reorientation of particles possessing a bare tumbling parameter $\lambda$ and shear susceptibility $\chi$.
Backflow is accounted for by balancing the change in angular momentum generated by the orientational collision operator with an angular momentum conserving term in the velocity collision operator~\cite{Noguchi2007}, the magnitude of which is governed by a rotational friction coefficient $\gamma_\text{R}$.

The N-MPCD method allows for simulations of two and three-dimensional nematic fluids, possessing isotropic-nematic phase transitions with annihilation of defects in 2D.
Subsequent analysis has shown that the N-MPCD method obeys linearized nematohydrodynamics under the assumption of isotropic viscosity and elasticity~\cite{hijar2019,hijar2019spontaneous,hijar2024particle}. 
All particles have the same mass $m$.
The N-MPCD particle mass $m$, thermal energy $\kbt$ and cell size $a$ set the simulation units, including units of time $t_0 = a\sqrt{m/\kbt}$.
The N-MPCD streaming time step is set to $\delta t = 0.1 t_0$ and the N-MPCD number density is $\rho = 20/a^3$ giving a Schmidt number of $\approx$ $375$~\cite{Ripoll2005}.
The N-MPCD parameters are chosen to be $\gamma_\text{R} = 0.01\kbt t_0$, $\lambda = 2$ and $\chi = 0.5$.
The nematic interaction constant $U_\text{NI} = 6\kbt$ is within the nematic phase~\cite{shendruk2015} but with an energy landscape that is low enough for the polymer to explore conformational space in computationally feasible times.

\begin{figure}
     \centering
     \includegraphics[width=0.99\linewidth]{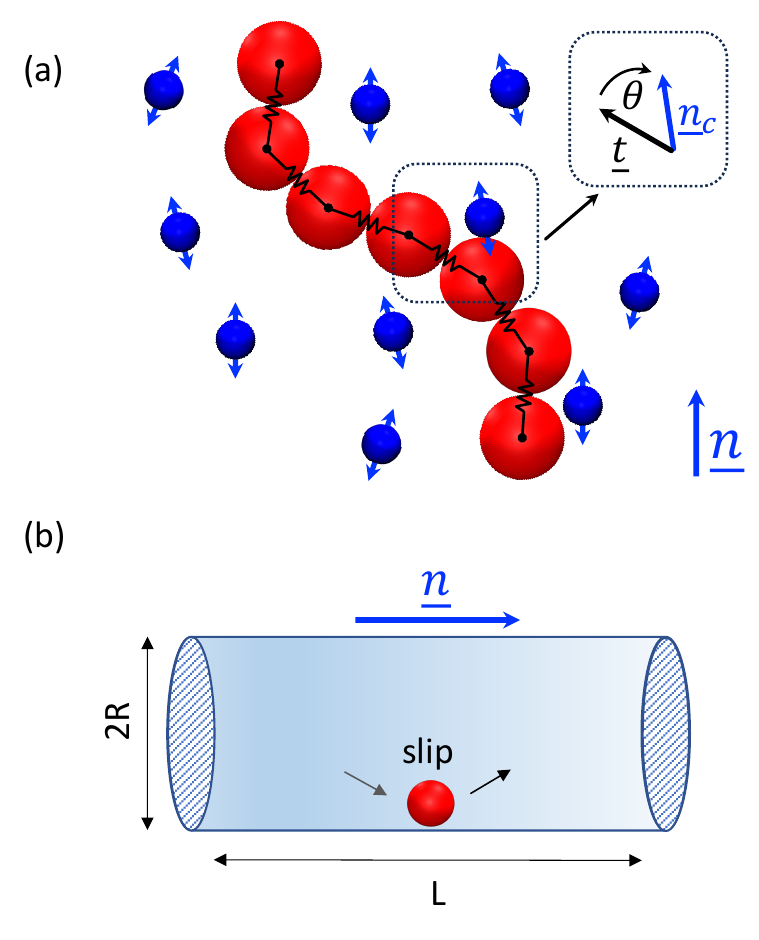}
     \caption{Simulation and boundary conditions. 
     (a) The orientational coupling potential between the polymer and nematic solvent is an angular harmonic potential (\eq{eq:Coupling}) between the backbone tangent $\vec  t$ and the local nematic director $\vecP{\vec n}{c}$. 
     (b) Cylindrical system of length $L = 30a$ and radius $R = 10a$. The director $\vec{n}$ is parallel to the long axis of the cylinder, with planar anchoring. The cylinder is impermeable with hydrodynamic slip conditions and periodic boundaries at both ends.}
     \label{fig:coupling}
\end{figure}
%

\subsection{Polymer Model}
\label{sec:MD}
The flexible polymer is simulated by molecular dynamics (MD)~\cite{hospital2015molecular} simulations. 
It is composed of $N$ beads $\Brac{j = 1,...,N}$ with mass $M$, which obey the equations of motion
\begin{align}
    \label{eq:EquationOfMotion}
    M\Ddot{\vecP{\vec r}{j}} = - \nabla U_j +
    \vecP{\vec \VCO}{j,c},
\end{align}
in which $U_j$ is the total potential of particle $j$ and $\vecP{\vec \VCO}{j,c}$ is the thermal and hydrodynamic drag forces due to including MD particle $j$ in the MPCD collision of cell $c$.
The beads interact via pair potentials which are composed of a bond $\vecP{U}{\text{B}}$, steric effects $\vecP{U}{\text{LJ}}$ and nematic coupling $\vecP{U}{\text{NPC}}$ terms.
We model freely jointed chains with no internal bending potential.

Beads are linearly connected by a finitely extensible nonlinear elastic bond potential~\cite{Grest1986,Kremer1990,Slater2009}
\begin{align}
    \label{eq:FENE}
    \vecP{U}{\text{B}} \Brac{\vecP{r}{jk}}
    = -\frac{k_\text{B}}{2} {r_0}^2 \ln 
    \Brac{ 1-\frac{\vecP{r}{jk}^2}{{r_0}^2}},
\end{align}
where $k_\text{B}$ is the bond strength, $r_0$ is the equilibrium length of the bonds and $\vecP{r}{jk} = \abs{\vecP{\vec r}{jk}}$ for $\vecP{\vec r}{jk} = \vecP{\vec r}{j} - \vecP{\vec r}{k} $ between monomers $j$ and $k$.
For bonds (\eq{eq:FENE}), $k=j-1$.
Excluded-volume interactions are taken into account by the purely repulsive Lennard-Jones potential~\cite{Weeks1971,Grest1986,Kremer1990,Slater2009}
\begin{equation}
    \label{eq:LJ}
    \vecP{U}{\text{LJ}}\Brac{\vecP{r}{jk}} = 4\epsilon
    \begin{cases}
         \Brac{\frac{\sigma}{\vecP{r}{jk}}}^{12} - \Brac{\frac{\sigma}{\vecP{r}{jk}}}^6
        + \frac{1}{4},
        & \vecP{r}{jk}<\sigma_{\text{CO}}\\
        0,
        & \vecP{r}{jk}>\sigma_{\text{CO}}.
    \end{cases}
\end{equation}
The energy $\epsilon$ sets the strength of the repulsive potential, $\sigma$ is the effective size of a bead and $\sigma_{\text{CO}} = \sqrt[6]{2}\sigma $ is the cutoff. 

Segments are coupled to their local nematic direction $\vecP{\vec n}{c}$ by a harmonic potential
\begin{equation}
    \label{eq:Coupling}
    \vecP{U}{\text{NPC}}\Brac{ \vecP{\vec t}{jk} ; \vecP{\vec n}{c} }
    = \frac{k}{2}
    \arccos^2 \Brac{ \vecP{\vec t}{jk} \cdot \vecP{\vec n}{c} }
    = \frac{1}{2}k {\theta}^2,
\end{equation}
where $\vecP{\vec t}{jk} = \Brac{ \vecP{\vec r}{j}-\vecP{\vec r}{k} }/\vecP{r}{jk}$ for $k=j-1$ is the tangent, $\theta$ is the angle between $\vecP{\vec t}{jk}$ and $\vecP{\vec n}{c}$, and $k$ the coupling constant (\fig{fig:coupling}a).
To balance the rotation of the segment, the N-MPCD mesogens within cell $c$ are subject to the torque
\begin{equation}
    \label{eq:Torque}
    \vecP{\vec \tau}{c} = \chi \Brac{ \vecP{\vec n}{c} \cdot \vecP{\vec \tau}{jk}}\Brac{ \vecP{\vec n}{c} \times \vecP{\vec \tau}{jk}},
\end{equation}
where $c$ is the cell that monomer $j$ resides within, and
$\vecP{\vec \tau}{jk}$
is the torque due to $\vecP{U}{\text{NPC}}$ on the segment connecting monomers $j$ and $k$.
The torque is parallel to $\vecP{\vec t}{jk}$, which assures that the local director feels an equal-but-opposite torque.
This torque changes the orientation of local nematic mesogens by
\begin{equation}
    \label{eq:TorsionalAngle}
    \delta \theta = \frac{\vecP{\vec \tau}{c}}{\gamma_\text{R}} \delta t,
\end{equation}
where $\gamma_\text{R}$ is the rotational friction coefficient.

In our simulations, parameters are set to $\epsilon = 1\kbt$, $\sigma = 1a$ and monomer mass $M =10m$ for all $j$.
The equilibrium bond length is $r_0 =1a$, with a bond strength $k_{\text{B}} =120\kbt/a^2$ and the coupling coefficients are varied from $k = 0$ to $20\kbt$.
The degree of polymerization is $N = 20$, giving a contour length $l = 19 b$, where $b = \Brac{0.89 \pm 0.02}a$ is the average bond length.
The MD algorithm time step is ${\delta t}_{\text{MD}} = 0.002t_0$, requiring $50$ MD iterations per N-MPCD iteration.  

\subsection{System Setup}

The simulations are conducted in a $3$D cylinder of length L $=30a$ and radius R$=10a$ (Fig. ~\ref{fig:coupling}b).
Strong planar anchoring at the cylinder surface assures the global liquid crystal orientation is parallel to the long axis of the cylinder ~\cite{head2024entangled}.
A perfect slip boundary condition on the impermeable cylinder wall is applied to velocity by reflecting the normal component of the velocity relative to the surface and leaving the tangential component unchanged.
Periodic boundaries cap both ends of the cylinder.
The fluid is initialized with Maxwell-Boltzmann distributed speeds and the director field parallel to the cylinder axis.
The polymer is initiated in the fully extended conformation on the center line of the cylinder, aligned with the global nematic direction and allowed to relax.
Data is recorded once the system reaches its steady state, which is identified via an iterative procedure.
In each iteration, the ensemble average is compared to the overall average calculated from all repetitions to find the first instance where it falls below this overall average.
This time is then used as the starting point for updating the overall average.
The process is repeated until successive updates of this reference time do not alter and stabilize, indicating that the system has achieved a steady state.
Twenty repeats for each set of parameters are simulated, each lasting $1.2-1.5\times 10^5 t_0$.


\section{Results}
First we present how conformational properties of the polymer are affected by different coupling parameters $k$. We then show how these configurations contribute to different diffusivity of the polymer. Finally, we quantify the dynamics of hairpins.    

\subsection{Conformations}
\label{sec:conformations}
The average shape of the polymers can be characterized by the gyration tensor
\begin{align}
    \label{eq:GyrationTensor}
    \tens G = \av{
    \Brac{ \vecP{\vec r}{i} - \vecP{\vec r}{\text{cm}} }
    \otimes
    \Brac{ \vecP{\vec r}{i} - \vecP{\vec r}{\text{cm}} }},
\end{align}
which measures the distribution of monomers around the center of mass, $\vecP{\vec r}{\text{cm}} = \av{\vecP{\vec r}{i}}$.
The average $\av{\cdot}$ is over the $N$ monomers.
The operation denoted by $\otimes$ is the outer product of vectors.
To understand how the nematic solvent impacts the conformation, we compare the values of $\tens G$ parallel and perpendicular to the global nematic orientation $\vec n$. 

\begin{figure}
    \centering
    \includegraphics[width=0.925\linewidth]{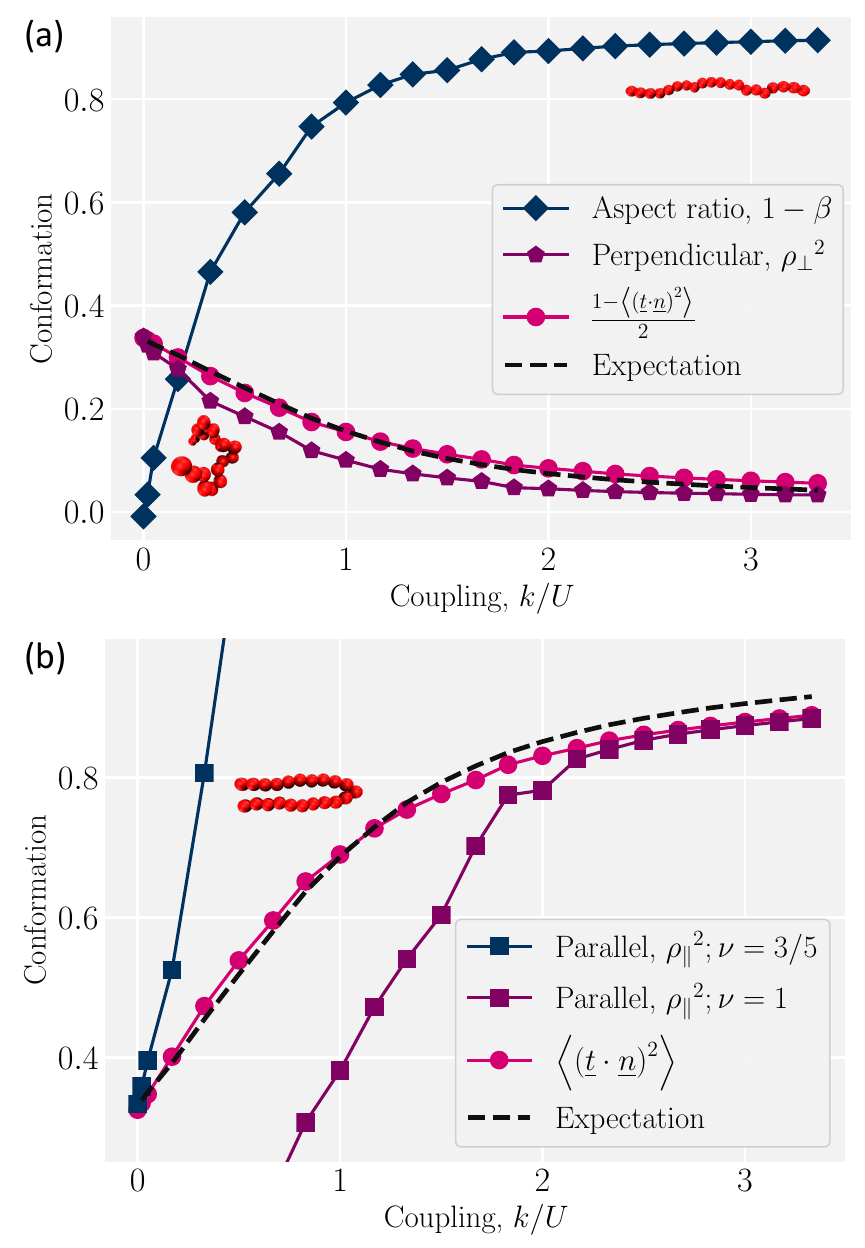}
    \caption{Coil-to-rod transition of polymer.
    (a) The aspect ratio is plotted as $1-\beta=1- {R_g}_\perp/{R_g}_\parallel$.
    The normalized perpendicular end-to-end distance is ${\rho_\perp}^2 ={\vecP{\vec R}{e,\perp}}^2/N^{2\nu}b^2$ with $\nu=3/5$ for self-avoiding random walk.
    The average of tangent perpendicular to the nematic direction
    $\Brac{1-\av{\Brac{ \underline t \cdot \underline n }^2}}/2$ 
    for simulations and the thermodynamic expectation value. 
    The bottom left and upper right snapshots illustrate typical coil-like and elongated conformations, respectively. 
    (b) The parallel end-to-end distance
    ${\vecP{R}{e,\parallel}}^2$ is normalized by different scalings, ${\rho_{\parallel}}^2 = {\vecP{\vec R}{e,\parallel}}^2/N^{2\nu}b^2$.
    At low coupling, $\nu = 3/5$ but $\nu = 1$ for a rod at high coupling.
    The average of tangent along the nematic direction $\av{\Brac{\vec t \cdot \vec n}^2}$ is compared to the thermodynamic expectation value.
    The snapshot shows a single hairpin.}
    \label{fig:conformation_k}
\end{figure}

Due to the strong anchoring, the global director lies along the axis of the cylinder. 
The parallel component of the gyration tensor is ${{R_g}_\parallel}^2= \vec n \hspace{1pt} \vec n : \tens G$ and perpendicular ${{R_g}_\perp}^2= \Brac{\Brac{\tens{1} - \vec n \hspace{1pt} \vec n}:\tens G}/2$, where the double dot product $:$ is the double contraction of tensors.
These form the semi-major and minor axes of an ellipsoid approximating the polymer.
These values give the aspect ratio $ \beta = {R_g}_\perp/{R_g}_\parallel$, which measures the degree to which the polymer is elongated.
When the polymer is weakly coupled to the nematic orientation $k/U \ll 1$, it conserves its symmetry, taking an isotropic shape with $1-\beta \approx 0$ (\fig{fig:conformation_k}a; diamonds).
As the coupling rises, the polymer becomes elongated in the nematic direction.
For high coupling $k/U \gg 1$,  it forms a rod-like conformation with $1-\beta \approx 1$ (\fig{fig:conformation_k}a; diamonds).
Coupling to the nematic orientation field elongates the polymer. 

To understand how the aspect ratio increases, let us assume a model in which the backbone of the polymer perpendicular to the global nematic director does a self-avoiding random walk.
This can be seen by considering the components of the end-to-end vector
${\vecP{\vec R}{e}} = \av{\vecP{\vec r}{N}-\vecP{\vec r}{1}}$.
The end-to-end distance of the perpendicular random walk
${\vecP{R}{e,\perp}} = \sqrt{{\vecP{\vec R}{e}} \cdot \Brac{\tens{1} - \vec n \hspace{1pt} \vec n} \cdot {\vecP{\vec R}{e}}}$
scales with the number of steps as $R_{e,\perp} \sim N^\nu$, where $\nu=3/5$. 
To compare this prediction to the simulations, we measure the perpendicular end-to-end distance
and normalize it as ${\rho_\perp} ={\vecP{\vec R}{e,\perp}}/N^{\nu}b$ for $\nu=3/5$ (\fig{fig:conformation_k}a; pentagons).
If the tangent $\vec t$ is decomposed into parallel ${\vec t}_\parallel = \vec n \hspace{1pt} \vec n \cdot \vec t$ and perpendicular ${\vec t}_\perp = \Brac{\tens{1} - \vec n \hspace{1pt} \vec n} \cdot \vec t$ components, the size of the steps in the perpendicular direction is
$b \sqrt{\av{\vec t_{\perp}^2}}$. 
Thus, we ideally expect ${\rho_\perp}^2 = \av{\vec t_{\perp}^2}$. 
At the low coupling limit $\Brac{k/U \ll 1}$, ${\rho_\perp}^2 = 1/3$ is expected for a coil-like random walk because the bond tangent vector $\vec t$ is a unit vector whose average in each direction is isotropic and so each component contributes $1/3$.
At high coupling $\Brac{k/U \gg 1}$, ${\rho_\perp}^2$ gradually goes to zero (\fig{fig:conformation_k}a; pentagons) as expected from the decreased degrees of freedom in the perpendicular direction and the suppression of $\av{\vec t_{\perp}^2}$.  

To further explore the role of the perpendicular fluctuations, we measure them directly from simulations as $\av{\vec t_{\perp}^2} = \Brac{1-\av{\Brac{ \underline t \cdot \underline n }^2 }}/2$ (\fig{fig:conformation_k}a; circles). 
The normalized end-to-end distance ${\rho_\perp}^2$ qualitatively agrees with $\av{\vec t_{\perp}^2}$.
However, there are quantitative differences, especially at intermediate couplings. 
We hypothesize that the difference is primarily due to the finite size of the polymer and build an analytical model for the large-$N$ thermodynamic limit. 

Assuming each segment fluctuates independently, the partition function --- the sum over all possible states --- of a single segment in the continuum limit is
\begin{equation}
    \label{eq:PartitionIntegral}
    Z_1 = \int_{0}^{2\pi} \,d\phi \int_{0}^{\pi} e^{-\frac{1}{2}\frac{k\hspace{1pt}b}{\kbt}\Brac{1- \Brac{\vec t \cdot \vec n}^2}} \sin\theta \,d\theta,
\end{equation}
which involves integrating over the azimuthal $\phi$ and polar $\theta$ angles.
The integral over the unit sphere gives
\begin{equation}
    \label{eq:PartitionFunction}
    Z_1 = e^{-\mathcal{K}} \sqrt{\frac{\pi}{\mathcal{K}}} \erfi \Brac{\sqrt{\mathcal{K}}}
\end{equation}
with the dimensionless number $\mathcal{K} = -k\hspace{1pt}b / \Brac{2\kbt}$.

We differentiate the free energy of one bond ($-\kbt \ln{Z_1}$) with respect to the coupling parameter $k$ to obtain the expectation value for its thermodynamic conjugate
\begin{align}
    \label{eq:AverageTangentPerp}
    1-\av{\Brac{\vec t \cdot \vec n}^2}
    = 1 + \frac{1}{2\mathcal{K}} - \frac{1}{\sqrt{\pi\mathcal{K}}} \frac{e^\mathcal{K}}{\erfi\Brac{\sqrt{\mathcal{K}}}},
\end{align}
which is equivalent to $2\av{\vec t_{\perp}^2}$ (\fig{fig:conformation_k}a; dashed line).
The thermodynamic expectation value agrees very well with the perpendicular tangent $\av{\vec t_{\perp}^2}$ but only qualitatively with the normalized end-to-end distance in the perpendicular direction ${\rho_\perp}^2$.
This leads us to conclude that, even though individual segments are in equilibrium, fluctuating about the global nematic director, the overall conformation is more complex than a simple sum of these fluctuations.
As we will show in \sctn{sec:HairpinFormation}
this is due to the existence of hairpins.

A similar line of argument follows for the parallel direction. 
Just like with the perpendicular component, the thermodynamic expectation value
\begin{align}
    \label{eq:AverageTangentPar}
    \av{\vec t_{\parallel}^2}
    = - \frac{1}{2\mathcal{K}} + \frac{1}{\sqrt{\pi\mathcal{K}}} \frac{e^\mathcal{K}}{\erfi\Brac{\sqrt{\mathcal{K}}}},
\end{align}
agrees well with simulation results for all coupling parameters (\fig{fig:conformation_k}b; dashed line and circles, respectively).  
The parallel component of the end-to-end vector
${\vecP{R}{e,\parallel}}^2 = {\vecP{\vec R}{e}} \cdot \vec n \hspace{1pt} \vec n \cdot {\vecP{\vec R}{e}}$ is more complicated. 
At low coupling $\Brac{k/U \ll 1}$, the polymer performs a self-avoiding random walk with the normalized parallel end-to-end distance ${\rho_{\parallel}}^2 = {\vecP{\vec R}{e,\parallel}}^2/N^{2\nu}b^2 = \av{\vec t_\parallel^2} = 1/3$ with $\nu=3/5$ (\fig{fig:conformation_k}b; blue squares). 
While the theory and simulations agree for the limit of very low coupling, the normalized end-to-end distance ${\rho_{\parallel}}^2$ quickly diverges and deviates from the parallel tangent. 
This is because the assumption of $\nu=3/5$ for an isotropic coil breaks down. 
Likewise, at high coupling $\Brac{k/U \gg 1}$, the polymer is a rod ${\rho_{\parallel}}^2 = {\vecP{\vec R}{e,\parallel}}^2/N^{2\nu}b^2 \rightarrow 1$ with $\nu = 1$ in this limit (\fig{fig:conformation_k}b; purple squares). 
The theory fails to reproduce the parallel extension $\rho_{\parallel}^2$ for intermediate coupling in the parallel direction due to crossing between scalings. 
This is due to the formation of hairpins --- sudden turns along the backbone of the polymer (\fig{fig:conformation_k}b, snapshot of the polymer).

\subsection{Hairpins}
\label{sec:HairpinFormation}
Hairpins arise in polymers suspended in nematic fluids because of the competition between conformational entropy and the energy due to being coupled to the nematic background~\cite{kamien1992theory}. 

For weak coupling between the polymer and the nematic $\Brac{k/U \ll 1}$, entropy maximization is dominant and polymers randomly explore conformational space (\fig{fig:conformation_k}a; low coupling snapshot). 
On the other hand for strong coupling ($k/U \gg 1$), the energy cost of deforming the surrounding liquid crystal is so high that the elongated conformations are preferred (\fig{fig:conformation_k}a; high coupling snapshot). 
However, at intermediate coupling neither contributions to the free energy are negligible. 
The polymer forms hairpin-like conformations to simultaneously satisfy the nematic symmetry and retain access to many conformational states.
The resulting hairpins are localized to sudden turns so that the energy cost of a hairpin is not substantial.
Because these hairpins can diffuse along the backbone of the polymer, the polymer can access many conformations with the same energy, allowing the entropy gain to compensate the energy cost. 

\begin{figure}
    \centering
    \includegraphics[width=0.99\linewidth]{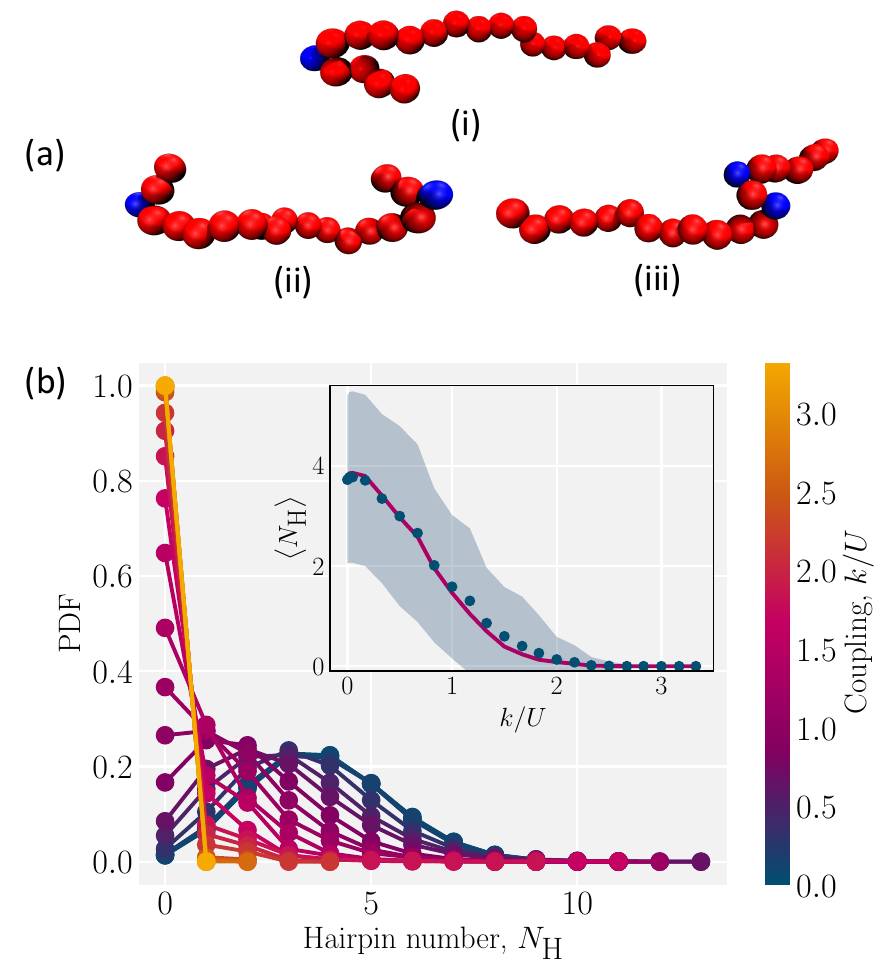}
    \caption{ Hairpin formation and distribution. 
    (a) Snapshots of different types of hairpin formation due to thermal fluctuations (i) single hairpin at one end (ii) Two hairpins at both ends (iii) Two hairpins (pair creation) away from ends. 
    (b) Probability distribution function (PDF) of number of hairpins $N_\text{H}$, for different coupling parameters $k$.
    (Inset) The measured average (dots) compared to the fit (solid line) of the PDF for each coupling to a Poisson distribution. The shaded region is the standard deviation.}
    \label{fig:HP_NO}
\end{figure}

Hairpins are formed by thermal fluctuations that are large enough to let the polymer overcome the elastic barrier of the nematic liquid crystal.
They form either as single hairpins forming at either ends (\fig{fig:HP_NO}a; i and ii) or as pairs at any point along the backbone (\fig{fig:HP_NO}a; iii).
By qualitative comparison of the results of simulations with experiments, we see that the single hairpin example (\fig{fig:HP_NO}a; i) is similar to top panel of \fig{fig:t4fd} from experiments and the two hairpins case (\fig{fig:HP_NO}a; ii) is similar to bottom panel of \fig{fig:t4fd}.

To identify hairpins, each monomer is given a ``hairpin score'', which measures the different features of hairpins. 
Monomers that have a sufficiently high score  are identified as hairpins (see \sctn{appendix:Hairpins} for details). 
Based on this identification, we measure the number of hairpins for different coupling.
For the low coupling $\Brac{k/U \ll 1}$, the probability distribution function (PDF) of the number of hairpins $N_\text{H}$ (\fig{fig:HP_NO}b) is wide.
The distribution has a maximum of 13 ``hairpins''. 
In this coil-like configuration, identifying these as hairpins is not particularly meaningful since the self-avoiding random walk has many sudden turns that are not related to the director orientation. 

As the coupling increases, the energy barrier rises and the likelihood of hairpin formation decreases.
This causes the PDFs to shift to lower values and narrow. 
The resulting lower average number of hairpins $\av{N_\text{H}}$ (\fig{fig:HP_NO}b; inset) has a narrower standard deviation than the weak coupling limit (\fig{fig:HP_NO}b; shaded area in inset).
The elastic nature of the nematic solvent dominates the thermal fluctuations when $k/U \gg 1$ and $\av{N_\text{H}} \rightarrow 0$ (\fig{fig:HP_NO}(b); inset). 
The number of hairpins are Poisson distributed, which is shown by comparing the measured average number of hairpins to the Poisson fit of the PDF (\fig{fig:HP_NO}b; inset). 
This suggests that hairpins form randomly and independently.

\begin{figure}
    \centering
    \includegraphics[width=\linewidth]{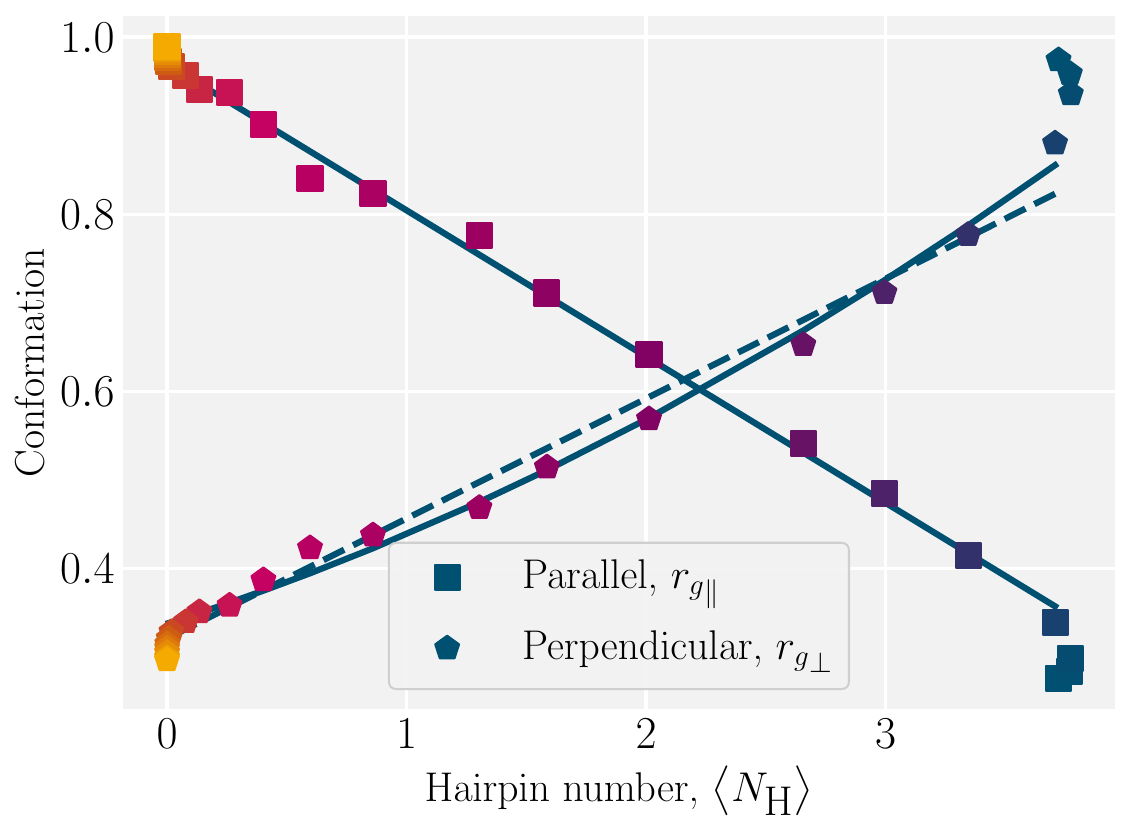}
    \caption{Conformation as a function of hairpin number $N_\text{H}$. The normalized parallel component of the radius of gyration is ${r_g}_\parallel = {R_g}_\parallel / R_\text{rod}$, where $R_\text{rod}$ is the radius of gyration for a rod. 
    The perpendicular component of gyration radius ${r_g}_\perp = {R_g}_\perp/R_\text{coil}$ is normalized by the perpendicular component of gyration radius for a self-avoiding polymer. 
    The colors correspond to the color bar in \fig{fig:HP_NO}. 
    At low hairpin number ($\av{N_\text{H}} \to 0 $), the conformation is rod-like with ${r_g}_{\parallel} \to  1$. At high hairpin number ($\av{N_\text{H}} 	\gtrsim 3 $), the conformation is a coil with ${r_g}_{\perp} \to 1$. The primary fits are plotted as solid lines, linear for the parallel component and quadratic for perpendicular component. A linear fit to the perpendicular component is plotted as a dashed line.
    }
    \label{fig:Conformation_HP}
\end{figure}

We previously demonstrated that the conformation is well predicted by theory in the low and high coupling limits, but hairpins strongly modify the observed conformations at intermediate couplings (\sctn{sec:conformations}). 
Having quantified the number of hairpins, the conformational properties at intermediate couplings can be explored as functions of number of hairpins. 
For this purpose, we return to the parallel and perpendicular components of gyration tensor. 
The normalized parallel component of gyration tensor is ${r_g}_\parallel={R_g}_\parallel/R_\text{rod}$, where $R_\text{rod} = N b/\sqrt{12}$ is the gyration radius for a rod with $N$ monomers connected by bonds of length $b$.
The normalized perpendicular component of gyration tensor is ${r_g}_\perp={R_g}_\perp/R_\text{coil}$, where $R_\text{coil}=bN^{3/5}\sqrt{25/528}$ is the perpendicular component of gyration radius for a self-avoiding polymer (\fig{fig:Conformation_HP}; symbols)~\cite{doi1986m}.
In the parallel direction, the gyration radius linearly decreases with average number of hairpins, ${r_{g}}_{\parallel} = \Brac{-0.165 \pm 0.003} \av{N_\text{H}} + \Brac{0.969 \pm 0.005}$ (\fig{fig:Conformation_HP}; solid line).
The linear dependence of ${r_g}_\parallel$ on $\av{N_\text{H}}$ highlights the significance of hairpins in the adopted shape of the polymers.

Similarly, the perpendicular component exhibits a linear dependence, ${r_{g}}_{\perp} = \Brac{0.135 \pm 0.005} \av{N_\text{H}} + \Brac{0.320 \pm 0.009}$ (\fig{fig:Conformation_HP}; dashed line).
While a linear fit is good, including a quadratic term ${r_{g}}_{\perp} = \Brac{0.337 \pm 0.007} + \Brac{0.086 \pm 0.012}\av{N_\text{H}} + \Brac{0.014 \pm 0.003}{\av{N_\text{H}}}^2$ further improves the agreement (\fig{fig:Conformation_HP}; solid line).
The goodness of these fits are assessed using the reduced chi-squared $\chi^2$.
The values of $\chi^2 = 0.33$ for the linear and $\chi^2 = 0.20$ for the quadratic fit indicate that both are acceptable but that including the quadratic term better represents the results.

%


\subsection{Dynamics}
\label{sec:dynamics}
\subsubsection{Center of mass dynamics}
The hairpins' impact on polymer dynamics can be seen by focusing on the polymer center-of-mass dynamics.
The dynamics of polymer center of mass is characterized by its mean squared displacement (MSD)
\begin{equation}
    \label{eq:MSD}
    \av{\delta \vecP{{\vec r}^2}{\text{cm}}} = 2dDt,
\end{equation}
where
$\delta \vecP{\vec r}{\text{cm}} =  \vecP{\vec r}{\mbox{\small{cm}}}\Brac{t+\delta t} - \vecP{\vec r}{\mbox{\small{cm}}}\Brac{t}$.
From equation~\ref{eq:MSD}, the isotropic diffusion coefficient $D$ is determined for the dimensionality $d$.
However, to understand how the nematic solvent affects the diffusion of the center of mass of the polymer, the MSD is decomposed into parallel and perpendicular components
\begin{equation}
    \label{eq:MSD-parallel}
         \av{ \delta \vecP{\vec r}{\text{cm}} \cdot \vec n \hspace{1pt} \vec n
         \cdot \delta \vecP{\vec r}{\text{cm}} }
        = 2dD_\parallel t
\end{equation}
\begin{equation}
    \label{eq:MSD-perp}
     \av{ \delta \vecP{\vec r}{\text{cm}} \cdot \Brac{\tens{1} - \vec n \hspace{1pt} \vec n}
     \cdot \delta \vecP{\vec r}{\text{cm}} }
    = 2dD_\perp t.
\end{equation}

\begin{figure}
    \centering
    \includegraphics[width=\linewidth]{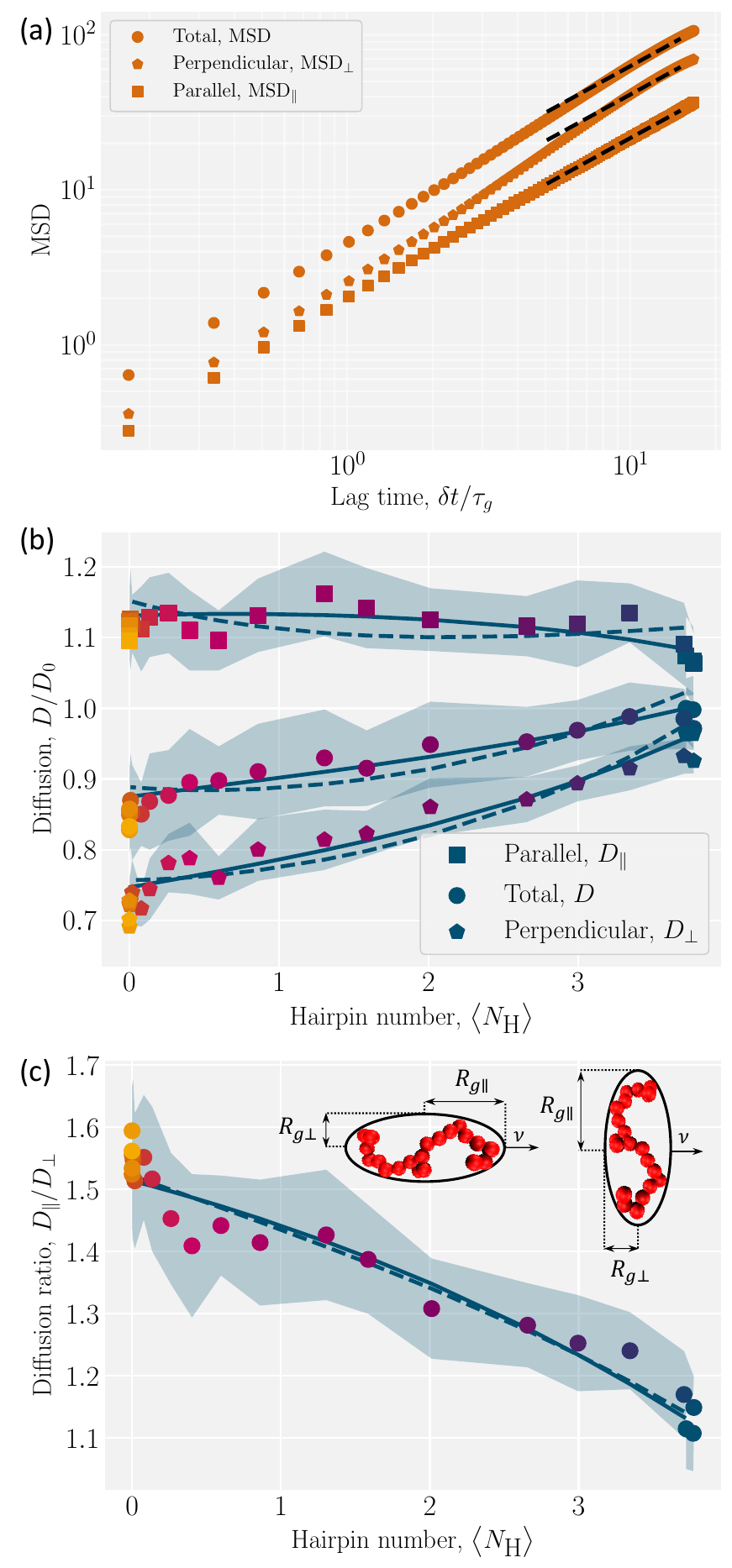}
    \caption{ Diffusion of polymer center of mass of as a function of hairpins number. 
    (a) An example MSD with its parallel ($\text{MSD}_\parallel$) and perpendicular ($\text{MSD}_\perp$) components for $k/U = 2.5$ with $\av{N_\text{H}} = 0$.
    The lag time is normalized by $\tau_g$, the time for the uncoupled ($k=0$) polymer to diffuse its size $R_g$.
    The dashed lines show the fits.
    (b) Total $D$, parallel $D_\parallel$ and perpendicular $D_\perp$ diffusivities are normalized by $D_0$, the uncoupled diffusion coefficient.
    (c) The ratio of parallel to perpendicular diffusivity. 
    (Inset) Ellipsoids of semi-major (${R_g}_\parallel$) and semi-minor (${R_g}_\perp$) axes, with different orientations to the direction of motion $\vec v$.
    In (b) and (c), the colors correspond to the color bar in \fig{fig:HP_NO} and solid lines show the ellipsoidal model, where the parallel gyration radius is fitted linearly and the perpendicular component quadratically. 
    The dashed lines are for the linear fit of the perpendicular component.
    Shaded regions shows the standard deviations.}
    \label{fig:CMdiffusion}
\end{figure}

An example MSD for coupling parameter $k/U = 2.5$ is plotted with its components in figure ~\ref{fig:CMdiffusion}a. 
Equations~\ref{eq:MSD-parallel} and~\ref{eq:MSD-perp} with $d=1$ and $d=2$ give the parallel $D_\parallel$ and the perpendicular $D_\perp$ diffusion coefficients, respectively. 
While the parallel diffusion stays unchanged as the average number of hairpins $\av{N_{\text{H}}}$ increases with decreasing coupling, the perpendicular and total diffusion increase as $\av{N_{\text{H}}}$ increases (\fig{fig:CMdiffusion}b).

To understand this anisotropy, consider two possible sources of diffusion anisotropy in liquid crystalline solvents.
The first is the anisotropy in the viscosity of the solvent. 
In nematic solvents viscosity is lower along the nematic director and
it has been shown that spheres diffuse anisotropically in a nematic solvent, with $D_\parallel /D_\perp \approx 1.6$ ~\cite{loudet2004stokes,stark2001stokes}.
The second source of the anisotropy is the shape of the solute.
For instance, for a rod-like inclusion, diffusion along the axis of the rod is less hindered than that in the direction perpendicular to the rod axis.
For an infinitely long thin rod in an isotropic solvent $D_\parallel /D_\perp = 2$ ~\cite{lauga2009hydrodynamics}. 
Since our nematic liquid crystal does not have anisotropic viscosity~\cite{hijar2019hydrodynamic}, the reason behind the measured anisotropic diffusion must be the anisotropy in the shape of the solute.
However, due to the finite length of the polymer, the ratio $D_\parallel /D_\perp$ converges to approximately 1.6, which is less than 2.

The diffusion coefficients are related to the drag coefficients through the Einstein relation as $\tens D = \kbt {\tens \zeta}^{-1}$ ~\cite{cappelezzo2007stokes} and so the anisotropy in the shape of the solute affects its diffusivity through anisotripic drag coefficients. The drag coefficients
\begin{equation}
    \label{eq:Drag}
    \tens \zeta \equiv 6 \pi \eta \mathcal{R} \tens K, 
\end{equation}
are related to the fluid viscosity $\eta$, the characteristic size of the solute $\mathcal{R}$ and its shape, accounted for by dimensionless resistance tensor $\tens K$
~\cite{yang2017bead}.
The resistance tensor $\tens K$ is known for various shapes, including prolate ellipsoids (\sctn{appendix:ellipse}).
Prolate ellipsoids are a reasonable first-order model for the elongated conformations of polymers in nematic fluids.
Approximating the polymer to first order as an ellipsoid gives the ratio of expected diffusivities for ellipsoids (denoted by superscript e) $D^e_\parallel/D^e_\perp$ as a function of only the aspect ratio $\beta$ (\sctn{appendix:ellipse}).
In an isotropic fluid when the shape is symmetric $\Brac{\beta=1}$, $D^e_\parallel = D^e_\perp$ and their ratio has its minimum value of unity.
For the rod-like limit $\Brac{\beta \ll 1}$ , the ratio rises to its maximum value of $D^e_\parallel/D^e_\perp = 2$ ~\cite{lauga2009hydrodynamics}.
We use this ellipsoidal model with semi-major axis ${R_g}_\parallel$ and semi-minor axis ${R_g}_\perp$ (\fig{fig:CMdiffusion}c; inset), to confirm that anisotropy in the diffusion rises only due to shape anisotropy.

However, the polymer is inside a cylinder and the cylinder wall affects polymer diffusion.
To account for the wall effect, correction factors are applied to the ellipsoidal expectation values~\cite{wakiya1957viscous},
\begin{align}
    \label{eq:DiffusionCorrection}
    D_\parallel &= \Brac{ 1-c_\parallel \frac{R_g}{R}} D^e_\parallel \\
    D_\perp &= \Brac{ 1-c_\perp \frac{R_g}{R}} D^e_\perp \nonumber ,
\end{align}
with the radius of the cylinder $R$ and $R_g = \sqrt{{R_g}_\parallel^2+2{R_g}_\perp^2}$ the gyration radius in the absence of coupling $\Brac{k = 0}$.
Applying these corrections to the values calculated using \eq{eq:Ellipsoid} reproduces the diffusion coefficients for the polymer embedded in a nematic inside the cylinder (\fig{fig:CMdiffusion}b-c; solid lines).
The correction parameters are fit to $c_\parallel = 1.20 \pm 0.01$ and $c_\perp = 1.35 \pm 0.02$.
The diffusion coefficients demonstrate good agreement with the simulation results.
The solid lines are calculated using the linear and quadratic fits for the parallel and perpendicular gyration radii, respectively
(\sctn{sec:HairpinFormation}).
While a linear fit to the perpendicular gyration radius was fairly accurate (\fig{fig:Conformation_HP}),
it causes considerable deviation from simulation values in the ellipsoidal model when compared to the quadratic fit (\fig{fig:CMdiffusion}b; solid lines vs. dashed lines).
In particular, the quadratic term does not curve up at high hairpin numbers, while the linear fit erroneously does.

The analysis shows that the shape is the primary factor contributing to anisotropic diffusion.
The variation in diffusion coefficients in different directions can be fully explained by the conformational changes resulting from coupling.
By increasing the coupling, number of hairpins decreases and polymer stretches and becomes less symmetric (\fig{fig:conformation_k}a).
The elongated polymers with few hairpins ($k/U \gg 1$) experience larger drag forces perpendicular to $\vec n$ which results in lower perpendicular diffusion coefficients, $\lim_{\av{N_{\text{H}}} \to 0} D_\perp  = \Brac{0.70 \pm 0.03}D_0$, with $D_0$ the diffusion coefficient of the center of mass of the polymer in the absence of coupling (\fig{fig:CMdiffusion}b).
On the other hand, as the coupling decreases the number of hairpins increases and the shape gets more symmetric with $D_\perp/D_0$ approaching unity, $\lim_{\av{N_{\text{H}}} \to \infty} D_\perp =  \Brac{0.96 \pm 0.02}D_0$ (\fig{fig:CMdiffusion}b).

In the parallel direction, diffusion starts at $\lim_{\av{N_{\text{H}}} \to \infty} D_\parallel  = \Brac{1.07 \pm 0.05}D_0$ for the limit of many hairpins and stays constant, only changing to $\lim_{\av{N_{\text{H}}} \to 0} D_\parallel  = \Brac{1.09 \pm 0.05}D_0$ for elongated polymers with no hairpin.
Perhaps this indicates that the parallel diffusion coefficient does not vary considerably with different numbers of hairpins.
However, the total diffusion coefficient $D$ gradually decreases down to $\lim_{\av{N_{\text{H}}} \to 0} D  = \Brac{0.83 \pm 0.05}D_0$ at the strongest coupling.
This should be expected since $D = \Brac{D_\parallel + 2D_\perp}/3$, which means that the decrease in $D_\perp$ mainly governs the total diffusion coefficient drop.
We must conclude that the mobility of the polymer is primarily controlled by the effect of its conformation on perpendicular diffusivity that is caused by its coupling to the nematic orientation. 

The ratio $D_\parallel/D_\perp$ shows a clearer comparison of the impact of coupling on diffusion coefficients in different directions (\fig{fig:CMdiffusion}c). 
In the low coupling $\Brac{k/U \ll 1}$, the ratio is $\lim_{\av{N_{\text{H}}} \to \infty} D_\parallel/D_\perp = 1.11 \pm 0.06$, which is close to the expected isotropic value $D_\parallel/D_\perp = 1$.
The deviation from unity is likely due to the cylinder wall that partially hinders the diffusion in the perpendicular direction.
The ratio for no hairpins ($k/U \gg 1$) is $\lim_{\av{N_{\text{H}}} \to 0} D\parallel/D_\perp = 1.56 \pm 0.1$.
This value is comparable to the ratio observed for a symmetric shape, such as a sphere, embedded in a nematic background with anisotropic viscosity, where $D\parallel/D_\perp = 1.6$ ~\cite{loudet2004stokes,stark2001stokes}.
This underscores the significance of the solute's shape in its diffusivity: changes to polymer conformation should not be neglected in estimating anisotropic diffusion since their contribution is comparable to the direct contribution of anisotropic viscosity.


\subsubsection{Hairpin dynamics}
\begin{figure}
    \centering
    \includegraphics[width=0.9\linewidth]{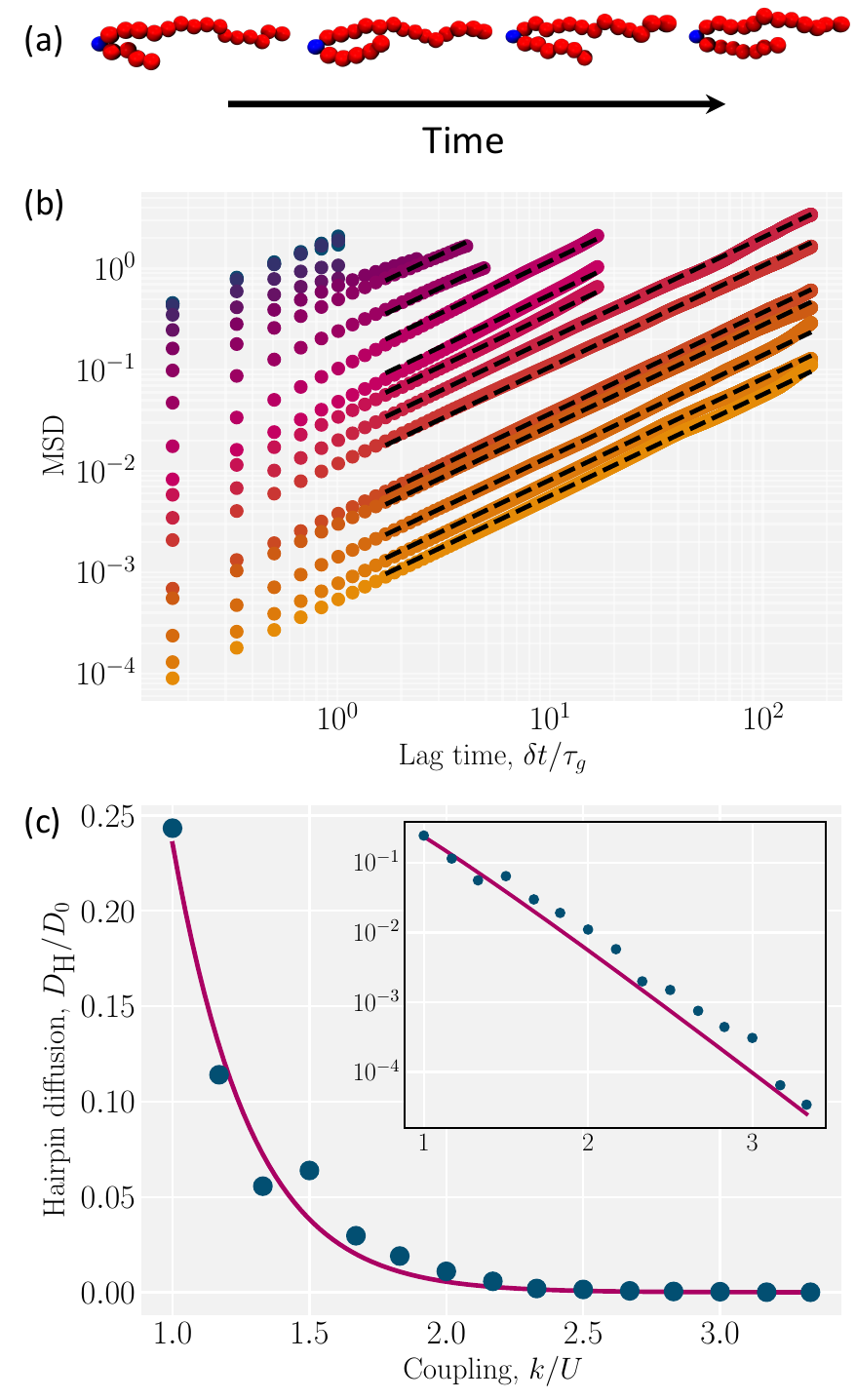}
    \caption{Hairpin diffusion. 
    (a) Snapshots of hairpin diffusion. The instantaneous position of the hairpin colored in blue. 
    (b) MSD examples for hairpin with various coupling parameter (the colors correspond to the color bar in \fig{fig:HP_NO}). The dashed lines represent the fit for diffusion coefficients. 
    (c) Diffusion of hairpin $D_\text{H}$ normalized by the uncoupled polymer center-of-mass diffusion $D_0$. 
    The solid line represents the exponential fit.
    }
     \label{fig:HP_diff}
\end{figure}

Hairpins are the only degree of freedom for an intermediately coupled polymer to explore its configuration space.
The movement of hairpins along the polymer backbone grants polymers access to different conformations through a diffusive hopping process.
We track hairpins over time (\fig{fig:HP_diff}a) and use their MSD along the backbone of the polymer to measure their diffusivity for different coupling parameters $k$ (\fig{fig:HP_diff}b). The diffusion coefficient of hairpins decreases exponentially as the coupling parameter increases  (\fig{fig:HP_diff}c; inset).  

To understand the exponential decay of the hairpin diffusion coefficient, consider the polymer backbone to be a one-dimensional lattice along which hairpins perform a hopping process.
Each time a hairpin hops, the bond connecting its current position to the next must rotate in the director field.
During this rotation, the polymer segment perturbs its surrounding liquid crystal orientation via the two-way coupling through $k$.
This suggests there is an energy cost to overcome, which can be understood through Kramer's model of Brownian motion across a barrier of height $E_b$ ~\cite{scherer2010theoretical}.
The probability current across the barrier sets the hopping rate ~\cite{scherer2010theoretical}
\begin{equation}
    \label{eq:HopingRate}
    \Gamma = \frac{1}{2\pi\zeta} \Brac{u^{\prime\prime}\Brac{x_a} u^{\prime\prime}\Brac{x_b}}^{\frac{1}{2}} \exp\Brac{-\frac{E_b}{\kbt}},
\end{equation}
where $\zeta$ is a drag coefficient of the hairpin, $u^{\prime\prime}\Brac{x_a}$ is the curvature of the energy well and $u^{\prime\prime}\Brac{x_b}$ is the curvature of the energy barrier over which it jumps.

For the hairpin hopping along the polymer backbone coupled to the nematic orientation,  $u^{\prime\prime}\Brac{x_a}= d^2\vecP{U}{\text{NPC}}/{d\theta}^2 = k$ and $u^{\prime\prime}\Brac{x_b}$ is estimated by the maximum change in the energy versus the minimum change in angle $u^{\prime\prime}\Brac{x_b} \equiv {\Delta u}_\text{max}/\Brac{{\Delta \theta}_\text{min}}^2$.
The maximum change in energy occurs when a bond transitions from being fully parallel to the nematic orientation to being fully perpendicular, ${\Delta u}_\text{max} = k\Brac{\pi/2}^2/2$.
For the small variation in angle, the arc length is $\Delta s \approx b{\Delta \theta}_\text{min} = v {\delta t}_\text{MD}$, which suggests ${\Delta \theta}_\text{min} = v {\delta t}_\text{MD}/b$.
Based on the equipartition theorem, the average energy for each degree of freedom is $\kbt/2$ which leads to $v = \sqrt{3\kbt/M}$ in 3D.  
Substituting these parameters in MPCD units into $u^{\prime\prime}\Brac{x_b} \approx {\Delta u}_\text{max}/\Brac{{{\Delta \theta}_\text{min}}}^2$ results in
$u^{\prime\prime}\Brac{x_a} \approx 90^2k$.
Finally, the barrier energy $E_b$ must be estimated. 
To proceed with this estimation, we assume a hairpin consists of two consecutive perpendicular bonds, each forming $\pi/4$ angle with the nematic orientation.
For the hairpin to hop to its next position, these bonds must move forward or backward along the polymer backbone.
The energy cost of this process is $k\Brac{\pi/2}^2/2-2\Brac{k \Brac{\pi/4}^2/2} = k\Brac{\pi/4}^2 = 0.62k$. 
Substituting these values into \eq{eq:HopingRate} and recognizing $D_H = b^2 \Gamma/2$ leads to the rough approximation $D_H/D_0 \approx 10 \frac{k}{U} \exp \Brac{-4\frac{k}{U}}$.
This approximation agrees well with the exponential fit of the hairpin diffusion coefficient $D_H/D_0 = \Brac{19.98\pm6.23} \frac{k}{U} \exp \Brac{-\Brac{4.44\pm0.29}\frac{k}{U}}$ (\fig{fig:HP_diff}c). 

We have shown that the hairpins move diffusively by hopping over local energy barriers.
In the strongly coupled limit $\Brac{k/U > 1}$ hairpins are topologically protected defects acting as singularities for the backbone tangent vector $\vec t$.
One can arbitrarily designate a hairpin that opens towards the $+\hat{x}$ direction (as in \fig{fig:HP_diff}a) to be a $+1/2$ hairpin ($1/2$ since the U-turn is 180$^\circ$ or half of $2\pi$) and a hairpin that opens in the $-\hat{x}$ direction a $-1/2$ hairpin.
As in active nematic systems~\cite{kozhukhov2022mesoscopic}, defects can arise through pair creation events and be removed through pair annihilation.
Additionally, individual hairpins spontaneously enter or leave the system from the polymer ends, explaining why odd $N_{\text{H}}$ are observed (\fig{fig:HP_diff}a and \fig{fig:HP_NO}-\ref{fig:CMdiffusion}).
%

\section*{Conclusion}
This study investigated the conformation and dynamics of a single flexible polymer suspended in a nematic liquid crystal background, where its backbone is coupled to the local nematic orientation.
This coupling introduces anisotropy to the polymer conformation and, if large enough, elongates the polymer.
In the weak and strong limits of coupling the extension of the polymer due to coupling can be described by the partition function for independent segments.
However, for intermediate coupling, the theory fails to reproduce the observed extension values due to formation of hairpin-like configurations along the polymer.
These hairpins act as topologically protected defects that minimize the internal energy by being mainly aligned with the liquid crystal and maximize entropy by moving along the polymer backbone.
We quantify the number of hairpins for each coupling strength to demonstrate that polymer conformational properties are characterized by the average number of hairpins.
The coupling leads to anisotropic diffusion of the polymer center of mass, mainly affecting the diffusivity in the perpendicular direction.
Since our nematic fluid has isotropic viscosity, this anisotropic diffusivity arises from asymmetric polymer shapes.
We employ the ellipsoidal model to incorporate this polymer shape anisotropy into its drag coefficient, which directly influences its diffusivity.
Our results show a good agreement with this model, confirming the shape-related anisotropy in polymer diffusion. 
We demonstrate that conformational effects of freely jointed polymers can be just as significant as viscosity anisotropy.
This demonstrates an independent mechanism to engineer dynamics of composite nematic/polymeric materials.
We further show how tuning the coupling strength can have a profound effect on hairpins dynamics, providing a potentially powerful mechanism for controlling the temporal dynamics of polymer configurations.

\section*{Conflicts of interest}
There are no conflicts to declare.

\section*{Acknowledgments}
This research has received funding (T.N.S.) from the European Research Council under the European Union’s Horizon 2020 research and innovation program (Grant Agreement No. 851196). 
A.R.K. is supported by the National Science Foundation of the United States, grant no. 2105113. We are grateful to Zvonimir Dogic and Christopher Ramirez for providing the sample of fd virus.
For the purpose of open access, the author has applied a Creative Commons Attribution (CC BY) license to any Author Accepted Manuscript version arising from this submission.

\bibliographystyle{unsrt}
\bibliography{rsc}

\begin{thebibliography}{10}

\bibitem{eder2018}
M.~Eder, S.~Amini, and P.~Fratzl.
\newblock Biological composites{\textemdash}complex structures for functional
  diversity.
\newblock {\em Science}, 362(6414):543--547, 2018.

\bibitem{Kikuchi2009}
N.~Kikuchi, A.~Ehrlicher, D.~Koch, J.A. K{\"a}s, S.~Ramaswamy, and M.~Rao.
\newblock Buckling, stiffening, and negative dissipation in the dynamics of a
  biopolymer in an active medium.
\newblock {\em Proceedings of the National Academy of Sciences},
  106(47):19776--19779, 2009.

\bibitem{secor2015filamentous}
P.R. Secor, J.M. Sweere, L.A. Michaels, A.V. Malkovskiy, D.~Lazzareschi,
  E.~Katznelson, J.~Rajadas, M.E. Birnbaum, A.~Arrigoni, K.R. Braun, et~al.
\newblock Filamentous bacteriophage promote biofilm assembly and function.
\newblock {\em Cell Host \& Microbe}, 18(5):549--559, 2015.

\bibitem{bidhendi2020fluorescence}
A.J. Bidhendi, Y.~Chebli, and A.~Geitmann.
\newblock Fluorescence visualization of cellulose and pectin in the primary
  plant cell wall.
\newblock {\em Journal of Microscopy}, 278(3):164--181, 2020.

\bibitem{oh2016}
D.X. Oh, Y.J. Cha, H.L. Nguyen, H.H. Je, Y.S. Jho, D.S. Hwang, and D.K. Yoon.
\newblock Chiral nematic self-assembly of minimally surface damaged chitin
  nanofibrils and its load bearing functions.
\newblock {\em Scientific Reports}, 6(1):1--6, 2016.

\bibitem{bansil2018biology}
R.~Bansil and B.S. Turner.
\newblock The biology of mucus: Composition, synthesis and organization.
\newblock {\em Advanced Drug Delivery Reviews}, 124:3--15, 2018.

\bibitem{witten2018selective}
J.~Witten, T.~Samad, and K.~Ribbeck.
\newblock Selective permeability of mucus barriers.
\newblock {\em Current Opinion in Biotechnology}, 52:124--133, 2018.

\bibitem{werlang2019}
C.~Werlang, G.~C{\'a}rcarmo-Oyarce, and K.~Ribbeck.
\newblock Engineering mucus to study and influence the microbiome.
\newblock {\em Nature Reviews Materials}, 4(2):134--145, 2019.

\bibitem{dogic2014}
Z.~Dogic, P.~Sharma, and M.J. Zakhary.
\newblock Hypercomplex liquid crystals.
\newblock {\em Annual Review of Condensed Matter Physics}, 5(1):137--157, 2014.

\bibitem{becerra2024conformational}
D.~Becerra, P.R. Jois, and L.M. Hall.
\newblock Conformational variability of intrinsically isotropic polymers with
  varying stiffness immersed in nematogenic solvents.
\newblock {\em Polymer}, 295:126774, 2024.

\bibitem{Barbara2007}
P.F. Barbara, W.S. Chang, S.~Link, G.D. Scholes, and A.~Yethiraj.
\newblock Structure and dynamics of conjugated polymers in liquid crystalline
  solvents.
\newblock {\em Annual Review of Physical Chemistry}, 58(1):565--584, 2007.

\bibitem{odijk1986}
T.~Odijk.
\newblock Theory of lyotropic polymer liquid crystals.
\newblock {\em Macromolecules}, 19(9):2313--2329, 1986.

\bibitem{dogic2004elongation}
Z.~Dogic, J.~Zhang, A.W.C. Lau, H.~Aranda-Espinoza, P.~Dalhaimer, D.E. Discher,
  P.A. Janmey, R.D. Kamien, T.C. Lubensky, and A.G. Yodh.
\newblock Elongation and fluctuations of semiflexible polymers in a nematic
  solvent.
\newblock {\em Physical Review Letters}, 92(12):125503, 2004.

\bibitem{lammi2004}
R.K. Lammi, K.P. Fritz, G.D. Scholes, and P.F. Barbara.
\newblock Ordering of single conjugated polymers in a nematic liquid crystal
  host.
\newblock {\em The Journal of Physical Chemistry B}, 108(15):4593--4596, 2004.

\bibitem{link2005}
S.~Link, D.~Hu, W.S. Chang, G.D. Scholes, and P.F. Barbara.
\newblock Nematic solvation of segmented polymer chains.
\newblock {\em Nano Letters}, 5(9):1757--1760, 2005.

\bibitem{turiv2013effect}
T.~Turiv, I.~Lazo, A.~Brodin, B.I. Lev, V.~Reiffenrath, V.G. Nazarenko, and
  O.D. Lavrentovich.
\newblock Effect of collective molecular reorientations on {B}rownian motion of
  colloids in nematic liquid crystal.
\newblock {\em Science}, 342(6164):1351--1354, 2013.

\bibitem{link2006}
S.~Link, W.S. Chang, A.~Yethiraj, and P.F. Barbara.
\newblock Anisotropic diffusion of elongated and aligned polymer chains in a
  nematic solvent.
\newblock {\em The Journal of Physical Chemistry B}, 110(40):19799--19803,
  2006.

\bibitem{lowen1999}
H.~L\"owen.
\newblock Anisotropic self-diffusion in colloidal nematic phases.
\newblock {\em Physical Review E}, 59:1989--1995, 1999.

\bibitem{egorov2016}
S.A. Egorov, A.~Milchev, and K.~Binder.
\newblock Anomalous fluctuations of nematic order in solutions of semiflexible
  polymers.
\newblock {\em Physical Review Letters}, 116:187801, 2016.

\bibitem{Note3}
Nematic solutions were obtained from Christopher Ramirez and Zvonimir Dogic at
  the University of California, Santa Barbara.

\bibitem{Dogic1997}
Z.~Dogic and S.~Fraden.
\newblock Smectic phase in a colloidal suspension of semiflexible virus
  particles.
\newblock {\em Physical Review Letters}, 78(12):2417, 1997.

\bibitem{tu2020direct}
M.Q. Tu, M.~Lee, R.M. Robertson-Anderson, and C.M. Schroeder.
\newblock Direct observation of ring polymer dynamics in the flow-gradient
  plane of shear flow.
\newblock {\em Macromolecules}, 53(21):9406--9419, 2020.

\bibitem{care2005}
C.M. Care and D.J. Cleaver.
\newblock Computer simulation of liquid crystals.
\newblock {\em Reports on Progress in Physics}, 68(11):2665--2700, 2005.

\bibitem{brini2013}
E.~Brini, E.A. Algaer, P.~Ganguly, C.~Li, F.~Rodríguez-Ropero, and N.F.A.
  van~der Vegt.
\newblock Systematic coarse-graining methods for soft matter simulations – a
  review.
\newblock {\em Soft Matter}, 9:2108--2119, 2013.

\bibitem{zannoni2018}
C.~Zannoni.
\newblock From idealised to predictive models of liquid crystals.
\newblock {\em Liquid Crystals}, 45(13-15):1880--1893, 2018.

\bibitem{allen2019}
M.P. Allen.
\newblock Molecular simulation of liquid crystals.
\newblock {\em Molecular Physics}, 117(18):2391--2417, 2019.

\bibitem{lebwohlLasher1972}
P.A. Lebwohl and G.~Lasher.
\newblock Nematic-liquid-crystal order---a {M}onte {C}arlo calculation.
\newblock {\em Physical Review A}, 6:426--429, 1972.

\bibitem{das2017}
R.~Das, M.~Kumar, and S.~Mishra.
\newblock Order-disorder transition in active nematic: A lattice model study.
\newblock {\em Scientific Reports}, 7(1):1--9, 2017.

\bibitem{shendruk2015}
T.N. Shendruk and J.M. Yeomans.
\newblock Multi-particle collision dynamics algorithm for nematic fluids.
\newblock {\em Soft Matter}, 11:5101--5110, 2015.

\bibitem{lee2017}
K.W. Lee and T.~Pöschel.
\newblock Electroconvection of pure nematic liquid crystals without free charge
  carriers.
\newblock {\em Soft Matter}, 13:8816--8823, 2017.

\bibitem{reyes2020defects}
D.~Reyes-Arango, J.~Quintana-H, J.C. Armas-P{\'e}rez, and H.~H{\'\i}jar.
\newblock Defects around nanocolloids in nematic solvents simulated by
  multi-particle collision dynamics.
\newblock {\em Physica A: Statistical Mechanics and its Applications},
  547:123862, 2020.

\bibitem{wamsler2024lock}
Karolina Wamsler, Louise~C Head, and Tyler~N Shendruk.
\newblock Lock-key microfluidics: simulating nematic colloid advection along
  wavy-walled channels.
\newblock {\em Soft Matter}, 20(19):3954--3970, 2024.

\bibitem{hijar2020dynamics}
H.~H{\'\i}jar.
\newblock Dynamics of defects around anisotropic particles in nematic liquid
  crystals under shear.
\newblock {\em Physical Review E}, 102(6):062705, 2020.

\bibitem{Noguchi2007}
H.~Noguchi, N.~Kikuchi, and G.~Gompper.
\newblock Particle-based mesoscale hydrodynamic techniques.
\newblock {\em Europhysics Letters ({EPL})}, 78(1):10005, mar 2007.

\bibitem{hijar2019}
H.~H{\'\i}jar.
\newblock Hydrodynamic correlations in isotropic fluids and liquid crystals
  simulated by multi-particle collision dynamics.
\newblock {\em Condensed Matter Physics}, 22(1):1--16, 2019.

\bibitem{hijar2019spontaneous}
H.~H{\'\i}jar, R.~Halver, and G.~Sutmann.
\newblock Spontaneous fluctuations in mesoscopic simulations of nematic liquid
  crystals.
\newblock {\em Fluctuation and Noise Letters}, 18(03):1950011, 2019.

\bibitem{hijar2024particle}
Humberto H{\'\i}jar and Apala Majumdar.
\newblock Particle-based and continuum models for confined nematics in two
  dimensions.
\newblock {\em Soft Matter}, 20(18):3755--3770, 2024.

\bibitem{Ripoll2005}
M.~Ripoll, K.~Mussawisade, R.G. Winkler, and G.~Gompper.
\newblock Dynamic regimes of fluids simulated by multiparticle-collision
  dynamics.
\newblock {\em Physical Review E}, 72:016701, 2005.

\bibitem{hospital2015molecular}
A.~Hospital, J.R. Go{\~n}i, M.~Orozco, and J.L. Gelp{\'\i}.
\newblock Molecular dynamics simulations: advances and applications.
\newblock {\em Advances and Applications in Bioinformatics and Chemistry},
  pages 37--47, 2015.

\bibitem{Grest1986}
G.S. Grest and K.~Kremer.
\newblock Molecular dynamics simulation for polymers in the presence of a heat
  bath.
\newblock {\em Physical Review A}, 33:3628--3631, May 1986.

\bibitem{Kremer1990}
K.~Kremer and G.S. Grest.
\newblock Dynamics of entangled linear polymer melts: A molecular‐dynamics
  simulation.
\newblock {\em The Journal of Chemical Physics}, 92(8):5057--5086, 1990.

\bibitem{Slater2009}
G.W. Slater, C.~Holm, M.V. Chubynsky, H.W. de~Haan, A.~Dubé, K.~Grass, O.A.
  Hickey, C.~Kingsburry, D.~Sean, T.N. Shendruk, and L.~Zhan.
\newblock Modeling the separation of macromolecules: {A} review of current
  computer simulation methods.
\newblock {\em Electrophoresis}, 30(5):792--818, 2009.

\bibitem{Weeks1971}
J.D. Weeks, D.~Chandler, and H.C. Andersen.
\newblock Role of repulsive forces in determining the equilibrium structure of
  simple liquids.
\newblock {\em The Journal of Chemical Physics}, 54(12):5237--5247, 1971.

\bibitem{head2024entangled}
L.C. Head, Y.A.G Fosado, D.~Marenduzzo, and T.N. Shendruk.
\newblock Entangled nematic disclinations using multi-particle collision
  dynamics.
\newblock {\em arXiv preprint arXiv:2404.09368}, 2024.

\bibitem{kamien1992theory}
R.D. Kamien, P.~Le~Doussal, and D.R. Nelson.
\newblock Theory of directed polymers.
\newblock {\em Physical Review A}, 45(12):8727, 1992.

\bibitem{doi1986m}
M.~Doi and S.F. Edwards.
\newblock {\em Theory of Polymer Dynamics}.
\newblock Oxford, 1988.

\bibitem{loudet2004stokes}
J.C. Loudet, P.~Hanusse, and P.~Poulin.
\newblock Stokes drag on a sphere in a nematic liquid crystal.
\newblock {\em Science}, 306(5701):1525--1525, 2004.

\bibitem{stark2001stokes}
H.~Stark and D.~Ventzki.
\newblock Stokes drag of spherical particles in a nematic environment at low
  {E}ricksen numbers.
\newblock {\em Physical Review E}, 64(3):031711, 2001.

\bibitem{lauga2009hydrodynamics}
E.~Lauga and T.R. Powers.
\newblock The hydrodynamics of swimming microorganisms.
\newblock {\em Reports on Progress in Physics}, 72(9):096601, 2009.

\bibitem{hijar2019hydrodynamic}
H~H{\'\i}jar.
\newblock Hydrodynamic correlations in isotropic fluids and liquid crystals
  simulated by multi-particle collision dynamics.
\newblock {\em Condensed Matter Physics}, 22(1):13601, 2019.

\bibitem{cappelezzo2007stokes}
M.~Cappelezzo, C.A. Capellari, S.H. Pezzin, and L.A.F. Coelho.
\newblock Stokes-{Einstein} relation for pure simple fluids.
\newblock {\em The Journal of Chemical Physics}, 126(22):224516, 2007.

\bibitem{yang2017bead}
K.~Yang, C.~Lu, X.~Zhao, and R.~Kawamura.
\newblock From bead to rod: Comparison of theories by measuring translational
  drag coefficients of micron-sized magnetic bead-chains in stokes flow.
\newblock {\em PLoS One}, 12(11):e0188015, 2017.

\bibitem{wakiya1957viscous}
S.~Wakiya.
\newblock Viscous flows past a spheroid.
\newblock {\em Journal of the Physical Society of Japan}, 12(10):1130--1141,
  1957.

\bibitem{scherer2010theoretical}
P.O.J. Scherer and S.F. Fischer.
\newblock {\em Theoretical Molecular Biophysics}.
\newblock Springer Science \& Business Media, 2010.

\bibitem{kozhukhov2022mesoscopic}
T.~Kozhukhov and T.N. Shendruk.
\newblock Mesoscopic simulations of active nematics.
\newblock {\em Science Advances}, 8(34):eabo5788, 2022.

\bibitem{probstein2005physicochemical}
R.F. Probstein.
\newblock {\em Physicochemical Hydrodynamics: An Introduction}.
\newblock John Wiley \& Sons, 2005.

\end{thebibliography}


\section*{Appendix}
\appendix
\section{Hairpin Identification}
\label{appendix:Hairpins}
To identify the hairpins, five hairpin factors considered. 
Each monomer is given a score $P_j = \frac{1}{5}\sum_{n=1}^{5}p_n$, where each $p_n$ measures one of the five hairpin features. 
Monomers with score  $P_j > P_\text{cutoff} = 0.5 $ are identified as hairpins. 
\paragraph*{Bend Aligns with Nematic.}
$p_1 = \abs{\vecP{\vec \kappa}{j} \cdot \vecP{\vec n}{c}}$ measures
the degree to which the unit curvature vector $\vecP{\vec \kappa}{j} = \Brac{\vecP{\vec t}{jj-1} - \vecP{\vec t}{j+1j}}/\abs{\vecP{\vec t}{jj-1} - \vecP{\vec t}{j+1j}}$ for monomer $j$ is aligned with the local nematic director.
If the segment is relatively straight, $\vecP{\vec \kappa}{j}$ is orthogonal to $\vecP{\vec n}{c}$. 
However, if there is a 180$^\circ$ turn, $\vecP{\vec \kappa}{j}$ is parallel to $\vecP{\vec n}{c}$ and this factor has the maximum value of 1.

\paragraph*{U-turn.}
$p_2 = \Brac{\vecP{\vec t}{jj-1} \cdot \vecP{\vec t}{j+1j} - 1}^2/4$
assesses the degree to which the segments after and before monomer $j$ are antiparallel .
For a U-turn conformation $p_2$ is close to 1; whereas, for self-avoiding random walks, this is rarely 1.

\paragraph*{Arm Length.}
$p_3$ measures the length of the two arms around a hairpin at index $j$. 
To determine the arm length, pairs of monomers a distance $\delta s$ away from $j$ are checked to see if they are antiparallel. 
The arm length is the maximum $\delta s$ for which they are antiparallel. 
It then normalizes the arm lengths by the maximum possible number arm length $l_\text{max} = \min\Brac{j, N-1-j}$.
Specifically, $p_3 = \frac{1}{l_\text{max}}\sum_{\delta s} \Theta\Brac{-\vecP{\vec t}{jj+\delta s}\cdot\vecP{\vec t}{jj-\delta s}}$, where $\Theta\Brac{\cdot}$ is the Heaviside function.
This value represents the relative length of hairpin arms.
 
\paragraph*{Relative Curvature.}
$p_4 = f\Brac{\abs{\vecP{\vec \kappa}{j}}/\av{\kappa}}$ measures the relative curvature at monomer $j$ in comparison with average curvature along the polymer.
The sigmoid function $f\Brac{x} = x/\sqrt{1+x^2}$ caps $p_4$ to 1. 
The average $\av{\cdot}$ is over the $N$ monomers

\paragraph*{Persistence.}
The last is a measure of persistence time of a hairpin at monomer $j$. To get this factor, we first check if the monomer $j$ was a hairpin in the previous time step.
If it was, the time possibility of being a hairpin now $p_\text{t}$ will be $1$. If it was not, then we check if it was an immediate neighbor to a hairpin.
If it was, it has $p_\text{t} = 1/2$.
Otherwise, $p_\text{t} = 0$.
Non-zero values of $p_\text{t}$ add to persistence time $\tau_j$ of being a hairpin at monomer $j$, $\tau_j\Brac{t+{\delta t}_{\text{MD}}} = \tau_j\Brac{t} + p_\text{t}$.
Conversely, when $p_\text{t} = 0$, the persistence time is halved $\tau_j\Brac{t+{\delta t}_{\text{MD}}} = \tau_j\Brac{t} / 2$.
Once again, the sigmoid function ensures values less than or equal to 1, by setting $p_5= f\Brac{\tau_j\Brac{t+{\delta t}_{\text{MD}}}
}$. 

\section{Ellipsoidal model}
\label{appendix:ellipse}
The diagonalized resistance tensor for an ellipsoid with an aspect ratio $\beta$ is
\begin{equation}
    \label{eq:ResistanceTensor}
    \tens K =
    \begin{bmatrix}
        K_\parallel & 0 \\
        0 & K_\perp
    \end{bmatrix},
\end{equation}
where
\begin{align}
    K_\parallel &= \frac{8}{3} \frac{1-\beta^2}{\Brac{ 2-\beta^2}\mathcal{S} - 2} \\
    K_\perp &= \frac{16}{3} \frac{1-\beta^2}{\Brac{ 2-3\beta^2}\mathcal{S} + 2} \nonumber\\
    \mathcal{S} &= 2 \Brac{ 1 - \beta^2 }^{-1/2} \ln \tr{ \frac{1+\Brac{1-\beta^2}^{1/2}}{\beta}}
\end{align}
for prolate ellipsoids $\beta<1$ ~\cite{probstein2005physicochemical}.
Thus, the expected parallel and perpendicular diffusion coefficients are  
\begin{align}
    \label{eq:Ellipsoid}
    D^{e}_\parallel &= \kbt {\zeta_\parallel}^{-1} = \frac{\kbt}{16 \pi \eta {r_g}_\parallel}\frac{\Brac{ 2-\beta^2}\mathcal{S} - 2}{1-\beta^2} 
    \\
    D^{e}_\perp &= \kbt {\zeta_\perp}^{-1} = \frac{\kbt}{32 \pi \eta {r_g}_\parallel}\frac{\Brac{ 2-3\beta^2}\mathcal{S} + 2}{1-\beta^2} \nonumber
\end{align}
and their ratio is
\begin{equation}
    \label{eq:DiffusionRatio}
    \frac{D^e_{\parallel}}{D^e_{\perp}} = 2 \frac{\Brac{1-\beta^2/2}\mathcal{S} -1}{\Brac{1-3\beta^2/2}\mathcal{S}+1}.
\end{equation}


\end{document}